\documentclass[letterpaper,twocolumn,prl,aps,superscriptaddress,amsmath,amssymb,floatfix]{revtex4-1}
\usepackage{mathptmx}
\usepackage[latin9]{inputenc}
\usepackage{graphicx}
\usepackage{epstopdf}
\usepackage{color}
\usepackage{verbatim}
\usepackage{float}
\usepackage{amsmath}
\usepackage{amssymb}
\usepackage{esint}
\usepackage{ulem}
\usepackage[T1]{fontenc}
\usepackage[unicode=true,
 bookmarks=true,bookmarksnumbered=false,bookmarksopen=false,
 breaklinks=false,pdfborder={0 0 1},backref=false,colorlinks=true]
 {hyperref}
\hypersetup{
 linkcolor=magenta,urlcolor=blue,citecolor=blue,pdfstartview={FitH},hyperfootnotes=false}

\makeatletter

\usepackage{textcomp}

\pdfpageheight\paperheight
\pdfpagewidth\paperwidth

\@ifundefined{textcolor}{}{%
 \definecolor{BLACK}{gray}{0}
 \definecolor{WHITE}{gray}{1}
 \definecolor{RED}{rgb}{1,0,0}
 \definecolor{GREEN}{rgb}{0,1,0}
 \definecolor{BLUE}{rgb}{0,0,1}
 \definecolor{CYAN}{cmyk}{1,0,0,0}
 \definecolor{MAGENTA}{cmyk}{0,1,0,0}
 \definecolor{YELLOW}{cmyk}{0,0,1,0}
}

\usepackage{xcolor}
\usepackage{soul}
\newcommand{\ket}[1]{\ensuremath{\left|#1\right\rangle}}

\definecolor{blue}{rgb}{0,0,1}
\definecolor{red}{rgb}{1,0,0}
\definecolor{green}{rgb}{0,1,0}

\usepackage{soul}

\makeatother

\begin{document}


\title{Transporting cold atoms towards a GaN-on-sapphire chip via an optical conveyor belt}
\author{Lei Xu}
\affiliation{CAS Key Laboratory of Quantum Information, University of Science and
Technology of China, Hefei 230026, China}
\affiliation{CAS Center for Excellence in Quantum Information and Quantum Physics,
University of Science and Technology of China, Hefei 230026, China}
\author{Ling-Xiao Wang}
\affiliation{CAS Key Laboratory of Quantum Information, University of Science and
Technology of China, Hefei 230026, China}
\affiliation{CAS Center for Excellence in Quantum Information and Quantum Physics,
University of Science and Technology of China, Hefei 230026, China}
\author{Guang-Jie Chen}
\affiliation{CAS Key Laboratory of Quantum Information, University of Science and
Technology of China, Hefei 230026, China}
\affiliation{CAS Center for Excellence in Quantum Information and Quantum Physics,
University of Science and Technology of China, Hefei 230026, China}
\author{Liang Chen}
\affiliation{CAS Key Laboratory of Quantum Information, University of Science and
Technology of China, Hefei 230026, China}
\affiliation{CAS Center for Excellence in Quantum Information and Quantum Physics,
University of Science and Technology of China, Hefei 230026, China}
\author{Yuan-Hao Yang}
\affiliation{CAS Key Laboratory of Quantum Information, University of Science and
Technology of China, Hefei 230026, China}
\affiliation{CAS Center for Excellence in Quantum Information and Quantum Physics,
University of Science and Technology of China, Hefei 230026, China}

\author{Xin-Biao Xu}
\affiliation{CAS Key Laboratory of Quantum Information, University of Science and
Technology of China, Hefei 230026, China}
\affiliation{CAS Center for Excellence in Quantum Information and Quantum Physics,
University of Science and Technology of China, Hefei 230026, China}

\author{Aiping Liu}
\affiliation{Institute of Quantum Information and Technology, Nanjing University of Posts and Telecommunications, Nanjing 210003, China}

\author{Chuan-Feng Li}
\affiliation{CAS Key Laboratory of Quantum Information, University of Science and
Technology of China, Hefei 230026, China}
\affiliation{CAS Center for Excellence in Quantum Information and Quantum Physics,
University of Science and Technology of China, Hefei 230026, China}

\author{Guang-Can Guo}
\affiliation{CAS Key Laboratory of Quantum Information, University of Science and
Technology of China, Hefei 230026, China}
\affiliation{CAS Center for Excellence in Quantum Information and Quantum Physics,
University of Science and Technology of China, Hefei 230026, China}
\affiliation{Hefei National Laboratory, University of Science and Technology of China, Hefei 230088, China}

\author{Chang-Ling Zou}
\email{clzou321@ustc.edu.cn}
\affiliation{CAS Key Laboratory of Quantum Information, University of Science and
Technology of China, Hefei 230026, China}
\affiliation{CAS Center for Excellence in Quantum Information and Quantum Physics,
University of Science and Technology of China, Hefei 230026, China}
\affiliation{Hefei National Laboratory, University of Science and Technology of China, Hefei 230088, China}

\author{Guo-Yong Xiang}
\email{gyxiang@ustc.edu.cn}
\affiliation{CAS Key Laboratory of Quantum Information, University of Science and
Technology of China, Hefei 230026, China}
\affiliation{CAS Center for Excellence in Quantum Information and Quantum Physics,
University of Science and Technology of China, Hefei 230026, China}
\affiliation{Hefei National Laboratory, University of Science and Technology of China, Hefei 230088, China}
\date{\today}

\begin{abstract}
Trapped atoms on photonic structures inspire many novel quantum devices for quantum information processing and quantum sensing. Here, we have demonstrated a hybrid photonic-atom chip platform based on a GaN-on-sapphire chip and the transport of an ensemble of atoms from free space towards the chip with an optical conveyor belt. The maximum transport efficiency of atoms is about $50\% $ with a transport distance of $500\,\mathrm{\mu m}$. Our results open up a new route toward the efficiently loading of cold atoms into the evanescent-field trap formed by the photonic integrated circuits, which promises strong and controllable interactions between single atoms and single photons.
\end{abstract}
\maketitle

\section{Introduction}
By introducing neutral atoms to integrated photonic devices, the hybrid photonic-atomic chip (PAC) has attracted extensive research in recent years~\cite{Chang2018,Luan2020,Beguin2020,Wang2022,Bouscal2023}. Benefiting from the strongly enhanced light-matter interactions due to the tightly optical field confinement at the wavelength and even subwavelength scale, PAC holds great potential in many quantum-based applications, such as quantum memory~\cite{lvovsky2009optical,gouraud2015demonstration}, novel quantum light sources~\cite{Pichler2017}, chiral quantum optics devices~\cite{Scheucher2016,Lodahl2017}, nodes of quantum networks~\cite{kimble2008quantum,tiecke2014nanophotonic,Muralidharan2017}, novel quantum optics phenomena with surface plasmons~\cite{stehle2014cooperative,stehle2011plasmonically}, many-body physics~\cite{Douglas2015,Gonzalez-Tudela2015}, and quantum sensing~\cite{zektzer2021nanoscale,sebbag2020chip}.

Early efforts on trapping cold neutral atoms near surface microstructures was firstly explored above current carrying microstructures ~\cite{Lin2004,Fortagh2002}, which can be tailored to create a variety of potential geometry and guiding schemes for cold atoms. Atoms are manipulated by the magnetic filed and the distances between the atoms and  microstructures surface can be reduced to as close as $0.5\,\mathrm{\mu m}$~\cite{Lin2004}.  With the advancement of fabrication techniques in photonic structures, manipulation of single atoms at wavelength and even subwavelength scale is possible with the tightly confined optical field confinement near the photonic structures. Many groundbreaking experimental results in coupling atoms to photonic structures have been achieved in various nanophotonic platforms~\cite{burgers2019clocked,thompson2013coupling,samutpraphoot2020strong,dhordjevic2021entanglement,kim2019trapping,zhou2021subwavelength,aoki2006observation,will2021coupling,vetsch2010optical,goban2012demonstration,gouraud2015demonstration,Corzo2019}. However, these researches move forward with some potential disadvantages. For instance, the platforms based on nanofibers~\cite{vetsch2010optical,goban2012demonstration,gouraud2015demonstration,Corzo2019} are suspended in vacuum, thus are potentially unstable and have poor thermal conductivity, which imposes limitations on the atom trap lifetime and atom coherence time~\cite{reitz2013coherence,hummer2019heating}. Besides, vacuum feedthrough for the coupling of light in and out of the nanofiber brings complexity in fiber alignment and assembly. Other platforms based on solid-state microcavities, such as microtoroid or bottle microresonators~\cite{aoki2006observation,Shomroni2014,Scheucher2016,will2021coupling}, unable to directly load laser cooled atoms into the evanescent-field trap~\cite{Barnett2000}, face difficulties in deterministic loading and trapping of cold atoms. Although the reported single atom-photon interaction time has been improved from only a few microseconds~\cite{aoki2006observation} to  $2\,\mathrm{ms}$~\cite{will2021coupling}, the further extension of the system to more photonic structures and more atoms is very challenging.

In order to achieve deterministic atom trapping on integrated photonic devices, important theoretical and experimental milestones have been reached with unsuspended waveguide and microring structures~\cite{burgers2019clocked,thompson2013coupling,samutpraphoot2020strong,dhordjevic2021entanglement,kim2019trapping,chang2019microring,zhou2021subwavelength,liu2022proposal}. Atoms are loaded into the  evanescent field of the photonic structures from free space with optical tweezers and optical conveyor belts. These methods exhibit a highly precise control of atomic motion near photonic structures, including photonic crystal waveguides~\cite{burgers2019clocked,thompson2013coupling,samutpraphoot2020strong,dhordjevic2021entanglement} and microring resonators~\cite{kim2019trapping,chang2019microring,zhou2021subwavelength}. Additionally, these demonstrations are compatible with on-chip integrated devices for cooling, transport, and trapping of cold atoms~\cite{liu2022multigrating,meng2015nanowaveguide,chen2022planar}.

In this work, we report on transporting free space cooled $^{87}$Rb atoms towards a GaN-on-sapphire chip~\cite{liu2022proposal} with an optical conveyor belt~\cite{burgers2019clocked,kuhr2001deterministic,nussmann2005submicron,dinardo2016technique}. The conveyor belt consists of two focused beams, both beams passing through the sapphire substrate perpendicularly. After careful spatial calibration of the beams and phase stabilization, our conveyor belt directly transports $ 10^4$ trapped atoms with a temperature around $40\,\mathrm{\mu K}$ toward the chip, without extra aberration of the focus beam from the sapphire substrate. The maximum transport efficiency of atoms is about $50\% $ with a transport distance of $500\,\mathrm{\mu m}$. It paves the way for the further implementation of stable atom trapping on the GaN-on-sapphire chip, promotes the realization of deterministic loading of atoms into the evanescent-field trap, which are promising for realizing the on-chip single-photon-level optical nonlinearity.

\begin{figure*}[t]
\center{\includegraphics[scale=1]{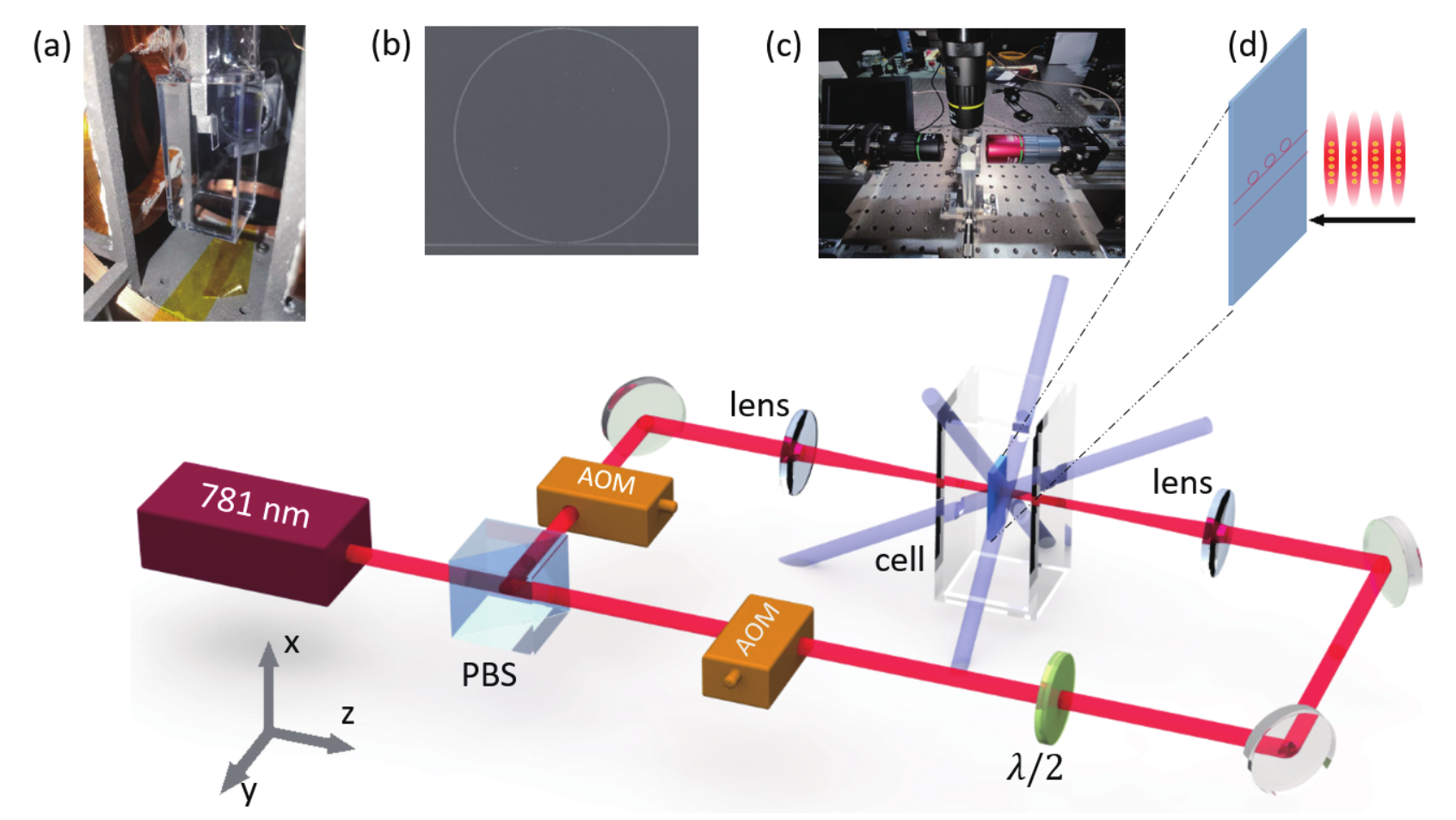}}
\caption{Schematic of our experimental system, with a GaN-on-Sapphire chip ($5\,\mathrm{mm} \times 10\,\mathrm{mm}$) platform placed inside the vacuum cell. Two pairs of cooling laser beams with crossing angle of 60 degree in X-Y plane go parallel to the chip surface, while the third pair passes through the crossing of the beams in X-Y plane and the chip surface at an angle of 60 degree to the chip surface. $\mathrm{781\, nm}$ dipole laser beams are splitted by a polarizing beam splitter (PBS), pass through the acousto-optic modulator (AOM) and focus onto the chip by a pair of lens with focus length $\mathrm{f = 100\, mm}$.  (a) Photograph of our PAC plaform in vacuum cell. (b) Picture of micro-fabricated microring resonator and the bus waveguides on top of the GaN-on-Sapphire chip taken by scanning electron microscope (SEM). (c) Photograph of our testing setup for the coupling to PAC. 	   (d) Schematic of optical conveyor belt for atom transport towards GaN-on-Sapphire chip.
}
\label{fig1} 
\end{figure*}

\section{Overview of the photonic-atom chip}

Figure~\ref{fig1} illustrates our experimental setup for studying the PAC, where a GaN-on-sapphire chip is placed inside the vacuum cell ($25\,\mathrm{mm}\times 25\,\mathrm{mm} \times 50\,\mathrm{mm}$). Figure~\ref{fig1}(a) shows a photograph of our PAC platform in a vacuum cell. We use low vapor pressure epoxy (Torr Seal) to stick the chip onto a metal holder while maintaining high vacuum and enduring the high temperature during the vacuum baking process. Half of the chip without fabrication is glued onto a 316-L stainless steel holder, and the remaining part of the chip is suspended in vacuum with waveguide and microring resonator structures fabricated on the surface. The metal holder is then connected to a CF35 vacuum cube, providing heat dissipation and stability for the chip.

Figure~\ref{fig1}(b) shows the scanning electron microscope (SEM) image of a fabricated microring resonator and the bus waveguides on top of the GaN-on-sapphire chip. The size of the sapphire substrate is $5\,\mathrm{mm} \times 10\,\mathrm{mm}$. The microring resonator is vacuum-cladded for direct interaction between atoms and the evanescent field of the confined modes, with a major radius of $60\,\mathrm{\mu m}$ and a cross-section of $700\,\mathrm{nm}\times420\,\mathrm{nm}$, and the optical modes of the microring are coupled to a bus waveguide through the evanescent field. Such microring resonators have been widely studied in photonics applications, due to their easy fabrication, high quality factor and small mode volume~\cite{Liu2022Review}. Here, we adapt the GaN-on-sapphire platform for PAC following our previous theoretical proposal~\cite{liu2022proposal}, as the system is more stable without suspended photonic structures. In addition, both GaN and sapphire are wide-band-gap materials that are transparent to ultrabroadband wavelengths (260 - 1590\,nm)~\cite{yu1997optical,muth1999absorption}, thus our chip is compatible with lasers working in the visible and near-visible wavelengths for many alkali and alkaline-earth atoms, allowing full optical access for cold rubidium atom cooling, trapping, transport and detection. The realization of the on-chip single-photon-level optical nonlinearity highly depends on the cooperativity parameter $C= \frac{3\lambda^2}{4\pi^2}\frac{Q}{V_m}$, where $\lambda = 780\,\mathrm{nm}$ is the wavelength of the D2 line of rubidium atoms. The cooperativity parameter is proportional to  the ratio of the quality factor $Q$  and the mode volume $V_m$  for the microring resonator .  For our microring resonator parameter,  the currently achieved quality factor $Q=3.75\times10^4$, mainly limited by the surface roughness.


As shown in Fig.~\ref{fig1}, our optical configurations of the experiments could be divided into three parts: (i) coupling to PAC, (ii) magneto-optical trap (MOT) , and (iii) optical conveyor belt. First, at both ends the GaN-on-sapphire chip, light is coupled in and out of the photonic chip through a high numerical aperture (N.A.) objectives. Figure~\ref{fig1}(c) shows the photograph of our testing setup for the coupling to PAC. A coupling efficiency of about $20\%$ for the GaN waveguide in Fig.~\ref{fig1}(b) for optical signals with $780\,\mathrm{nm}$ wavelength can be achieved with commercial $N.A.=0.4$ objectives. 

The cold $^{87}$Rb atoms are then prepared through a standard six-beam magneto-optical trap~\cite{Metcalf1999}. The glass cell is connected to a mini cube and a $30\,\mathrm{L/s}$ ion pump, resulting in a pressure of $10^{-9}$ mbar measured by the ion-pump current. Three pairs of cooling laser beams are generated from a $780\,\mathrm{nm}$ laser, with the power of each beam being about $150\,\mathrm{\mu W}$ and the beam waist being $1\,\mathrm{mm}$. The cooling laser beam is red detuned by $8\,\mathrm{MHz}$ from the $\mathrm{\ket{F = 2}\,\rightarrow\ket{F' = 3}}$ D2 cycling transition. Additionally, $80\,\mathrm{\mu W}$ of repump laser beams overlap with one of the cooling laser beams.  The beams intersect at one point about $1\,\mathrm{mm}$ above the surface of the chip, with additional anti-Helmholtz coils aligned with the point providing a magnetic field gradient up to $10\,\mathrm{G/cm}$. To align with our PAC, two pairs of cooling laser beams with a crossing angle of 60 degree in the X-Y plane go parallel to the chip surface, while the third pair passes through the crossing of the beams in the X-Y plane and the chip surface at an angle of 60 degree to the chip surface. Although MOT beam pairs are not oriented orthogonal to each other, 3D atom confinement is still achieved as components of each beam are projected along the orthogonal axis. 

\begin{figure}[b]
\center{\includegraphics[scale=1]{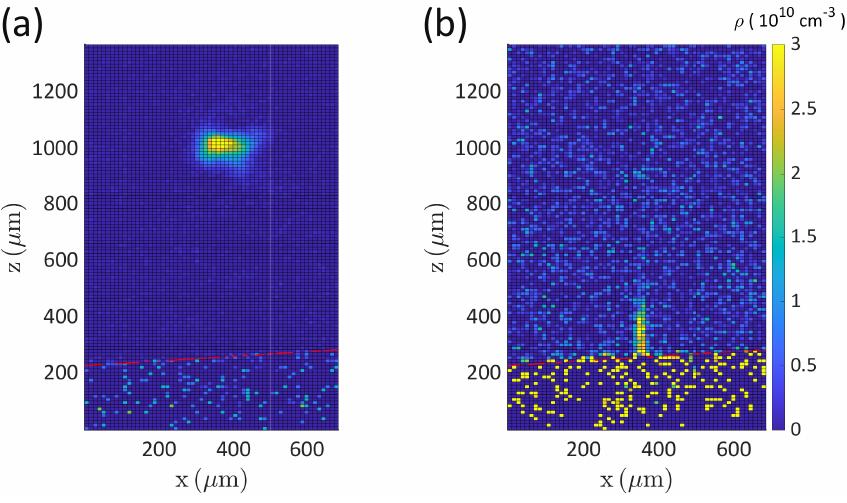}}
\caption{ (a) Single shot absorption images of MOT atoms $800\,\mathrm{\mu m}$ away from the chip surface. The absorption imaging is performed with camera of $2048\times1080$ resolution and a pixel size of $4.7\,\mathrm{\mu m}$. The imaging path to the camera has a magnification of 0.3. The red line denotes the interface between free space and photonic chip. (b) Single shot absorption images of atoms transported toward the chip surface with the optical conveyor belt.
 }
\label{fig2}
\end{figure}

Following the 3D MOT procedure, the temperature of the atom ensemble around $40\,\mathrm{\mu K}$ is finally achieved by a polarization-gradient cooling (PGC) process.  With a duration of $2\,\mathrm{ms}$ for the PGC, the cooling laser beams detuning is ramped down to $-48\,\mathrm{MHz}$ from the cycling transition. Figure~\ref{fig2}(a) shows a density contour plot of the atom ensemble, which is deduced from a single shot free space absorption image\cite{ketterle1999}. The cold atoms are about $800\,\mathrm{\mu m}$ away from the chip surface, and the atom number density is about $3\times 10^{10}\,\mathrm{cm^{-3}}$ with an atom cloud radius about $190\,\mathrm{\mu m}$. The distance between the atom cloud and the chip surface can be adjusted from $300\,\mathrm{\mu m}$ to $1000\,\mathrm{\mu m}$ by adjusting the offset coil and the alignment of the cooling beams. However, in close proximity to the chip, the density and shape of the atomic cloud is altered due to surface reflections, which is consistent with previous observations\cite{Huet2012}.

\begin{figure}[t]
\center{\includegraphics[scale=1]{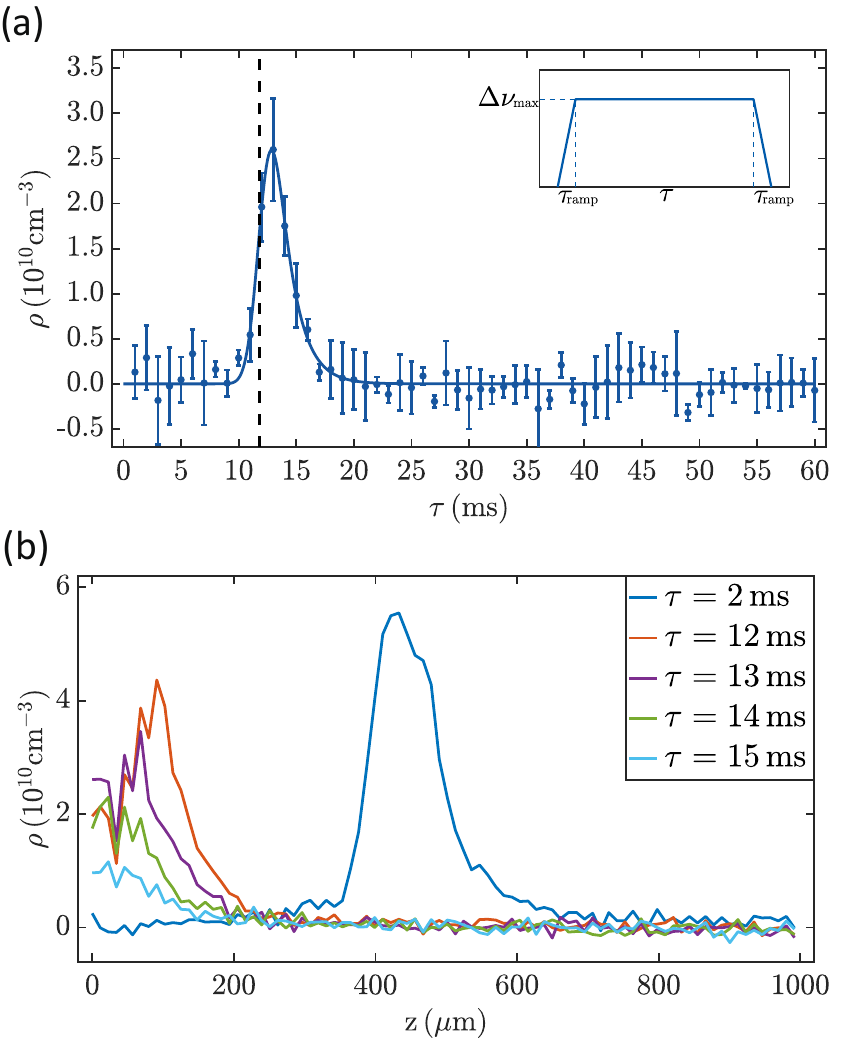}}
\caption{(a) Atomic density $\rho$ on chip surface with different holding times $\tau$.  We measure the trajectories of moving atoms in atom conveyor belt by taking a sequence of images with incremental holding times and record the atomic density on chip surface (averaged 10 times) with $\Delta\nu\mathrm{_{max}}=104\,\mathrm{kHz}$ and a maximum transport distance $d=500\,\mathrm{\mu m}$. We observe a clear peak at $\mathrm{\tau = 13\,ms}$, indicating a peak atom flux accumulating on the chip surface.  The error bar comes from three continuous measurement sequences. Solid curve is the fit result from our theoretical model of atomic accumulation dynamics on the chip surface (see details in the text). Inset is the laser detuning profile, the rising and descending time of the profile is set 1ms, and the holding time is $\tau$. (b) Atomic density distribution along the conveyor belt axis for different holding times.
}
\label{fig3}
\end{figure}

\section{Transporting atoms towards the chip}

The optical conveyor belt is realized by an optical dipole trap, which consists of two linearly polarized counter-propagating Gaussian beams with beam waist $w_0 = 20\,\mathrm{\mu m}$, and the waist is located in the middle of the MOT atom cloud and the chip surface. The intensities of both beams are equal, and their optical frequencies are different by detuning
$\Delta\nu$. The two counterpropagating dipole laser beams come from a single laser, with $\lambda = 781\,\mathrm{nm}$ corresponding to a frequency $2.3\,\mathrm{THz}$ red detuned to the $^{87}$Rb atom D2 transition. The laser is split into two paths and then passed through a double-pass 80 MHz acousto-optic modulator (AOM), with $\Delta\nu$ controllable via the RF signals applied to the AOMs. Therefore, the two beams generate a spatially and temporally varying trap potential $U(z,t) = U_0\cos^2(\frac{2\pi\Delta\nu}{2}t-\frac{2\pi}{\lambda}z)$, where $U_0$ is the local trap depth and $z$ is the position of atoms along the beam axis. Trapped atoms are confined in the lattice antinodes and moved with the temporally varying trap potential (see Fig.~\ref{fig1}(d)).

In order to transport atoms towards the surface of the PAC, we first load approximately $10^4$ atoms into a standing-wave dipole trap with $\Delta\nu=0$ by overlapping both dipole beams with the 3D MOT for $150\,\mathrm{ms}$. Here, each beam has a power of $9\,\mathrm{mW}$, which corresponds to a trap depth of about $1.3\,\mathrm{mK}$. After the loading process, atoms are trapped in a series of lattice antinodes along the beam axis, with an axial distribution of about $150\,\mathrm{\mu m}$ limited by the size of the atom cloud. Then, a frequency chirping sequence of $\Delta\nu$ (see the inset of Fig.~\ref{fig3}(a)), which is achieved by the sweep mode of the signal generator, is sent to one of the AOMs to create a moving optical conveyor belt, and the antinodes move at a velocity~\cite{kuhr2001deterministic}
\begin{equation}\label{eq1}
    v=\frac{1}{2}\lambda\Delta\nu.
\end{equation}
To verify the transport of atoms towards the chip surface, we take an absorption image of cold atoms in the optical conveyor belt after the transport process. Figure~\ref{fig2}(b) shows the results indicating the ensemble of atoms near the chip surface (the dashed red line). Compared with the image of the atom cloud prepared by MOT, our conveyor belt has successfully delivered atoms towards the chip. Then, the transport of atoms in the optical conveyor belt is systematically investigated. Through a sequence of $\Delta\nu$, as illustrated by the inset of Fig.~\ref{fig3}(a), we could transport the atoms over a certain distance by ramping up $\Delta \nu$ in $\tau_{\mathrm{ramp}}=1\,\mathrm{ms}$ to $\Delta\nu\mathrm{_{max}}$, holding the detuning for a duration of $\tau$, and then ramping down $\Delta \nu$ in $1\,\mathrm{ms}$. The distance $\Delta z =\frac{1}{2}\lambda\Delta\nu\mathrm{_{max}}(\tau_{\mathrm{ramp}}+\tau)$. Figure~\ref{fig3}(a) summarizes the measured atomic density on the chip surface for different hold times $\tau$, with $\Delta\nu\mathrm{_{max}}=104\,\mathrm{kHz}$ and a maximum transport distance $d=500\,\mathrm{\mu m}$ limited by the block of the chip. We observed a clear atomic density peak of $2.5\times10^{10}$ $\mathrm{cm^{-3}}$ when $\tau =13\,\mathrm{ms}$, which almostly agrees with the calculated time (dashed vertical line) for 
transporting atoms from the center of the MOT to the chip surface.

The accumulation of atomic density $\rho$ on the chip surface can be described by a simple rate equation
\begin{equation}\label{eq2}
	\frac{d\rho}{dt} =J(t)-\Gamma\rho,
\end{equation}
where $J(t)$ describes the atomic flux to the chip surface by the optical conveyor belt and $\Gamma$ is the linear atomic loss coefficient. Here, $\Gamma$ is mainly attributed to the atom collision and absorption loss on the chip surface, heating by the optical dipole trap, and the vacuum gas collisions. Since the atomic density is relatively low, the nonlinear atomic loss due to atomic collisions is neglected. According to the atomic cloud shape, we made the assumption that atomic flux is a Gaussian function
\begin{equation}\label{eq3}
    J(t)=J_0 e^{-(\frac{t-\tau_0}{\sigma_0})^2},
\end{equation}
where $J_0$ is the maximum atomic flux density, $\tau_0$ is the time of atomic peak flux arriving at the chip surface without the block of the chip, and $\sigma_0$ describes the width of the Gaussian function in the time domain. By the above rate equation, we fitted the atomic density (solid line) in Fig.~\ref{fig3}(a), which agrees excellently with the experimental results. 

Further investigations of the influence of the chip surface on the transport of atoms are shown in Fig.~\ref{fig3}(b), where the atomic density distribution along the atom conveyor belt axis for different holding times $\tau$ is plotted. The origin of the $z$-axis is set to the chip surface. The parameters of the conveyor belt are the same as in Fig.~\ref{fig3}(a). We found that when atoms are close to the chip surface within $100\,\mathrm{\mu m}$, the peak height of atomic density distribution along the atom conveyor belt axis decreases, which indicates an increased atom loss rate. 
  
\begin{figure}[t]
\center{\includegraphics[scale=1]{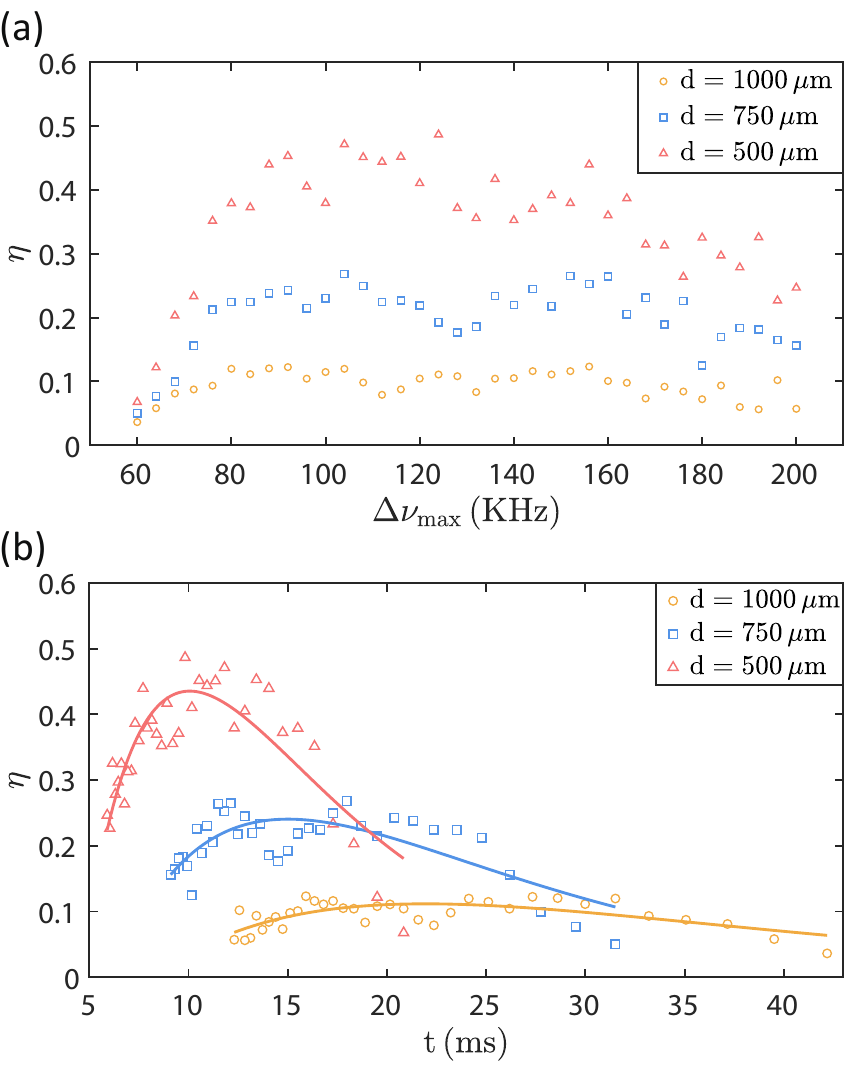}}
\caption{(a) Atom transport efficiency $\eta$ with maximum frequency difference $\mathrm{\Delta \nu\mathrm{_{max}}}$ for different transport distance $d$. The transport efficiency is defined as the ratio of the maximum atomic density on the chip surface to the initial atomic density in the standing dipole trap.  The transport efficiency reaches the optimum when $\Delta \nu\mathrm{_{max}}$ is $100-160\,\mathrm{kHz}$. (b) Atom transport efficiency $\eta$ with transport time $t$ for different transport distance $d$. Higher transport efficiency and shorter transport distance are achieved with shorter transport distance.
}
\label{fig4}
\end{figure}

For future PAC experiments, a high density of atoms on the surface of the GaN-on-sapphire chip is desired. Therefore, we experimentally varied the location of the MOT center, and optimized the $\Delta\nu\mathrm{_{max}}$ for atom transportation. The results for maximum transport distance $d = 1000\,\mathrm{\mu m}$,  $750\,\mathrm{\mu m}$, and $500\,\mathrm{\mu m}$ are summarized in Fig.~\ref{fig4}, with the atomic density on the chip surface calculated from the fitting results, as shown by the solid line of Fig.~\ref{fig3}(a). The transport efficiency $\eta$ is defined as the ratio of the maximum atomic density on the chip surface to the initial atomic density in the standing wave dipole trap. Comparing the different $\Delta \nu\mathrm{_{max}}$ in Fig.~\ref{fig4}(a), the transport efficiency reaches the optimum when $\Delta \nu\mathrm{_{max}}$ is $100-160\,\mathrm{kHz}$, and the transport efficiency drops when $\Delta \nu\mathrm{_{max}}$ is further increased. In particular, the efficiency dramatically decreases when $\Delta \nu\mathrm{_{max}}$ is less than $80\,\mathrm{kHz}$. The dependence of $\eta$ on $\Delta \nu\mathrm{_{max}}$ might be attributed to two different reasons. If $\Delta \nu\mathrm{_{max}}$ is too large, the acceleration and deceleration process of the conveyor belt might induce a significant loss of atoms. Such phenomena have been studied in many other experimental works in detail~\cite{Schrader2001,Hickman2020}. In contrast, if $\Delta \nu\mathrm{_{max}}$ is too small, the required $\tau$ is too large, and the atomic density is limited by the intrinsic atomic losses in the dipole trap. In practical cases, we aim to obtain higher transport efficiency along with less transport time. In  Fig.~\ref{fig4}(b), we present the transport efficiency with corresponding transport time for different transport parameter. Shortening the transport distance can improve both the transport efficiency and transport time, and a maximum transport efficiency close to $50\,\%$ is achieved for a maximum transport distance of $d = 500\,\mathrm{\mu m}$. We fit our results with an empirical equation of the form
\begin{equation}\label{eq4}
    \eta = e^{-at}(bt+c)
\end{equation}
to describe the relation between transport efficiency with corresponding transport time. The exponential decay accounts for the intrinsic loss irrelevant to the transport velocity, while the linear term accounts for the improvement of transport efficiency with slow transport velocity. It is anticipated that the transport efficiency can be improved by reducing the transport distance. However, the disturbing of the MOT when it is in the proximity of the chip surface prevent us from achieving a shorter distance.

\begin{figure}[t]
 \center{\includegraphics[scale=1]{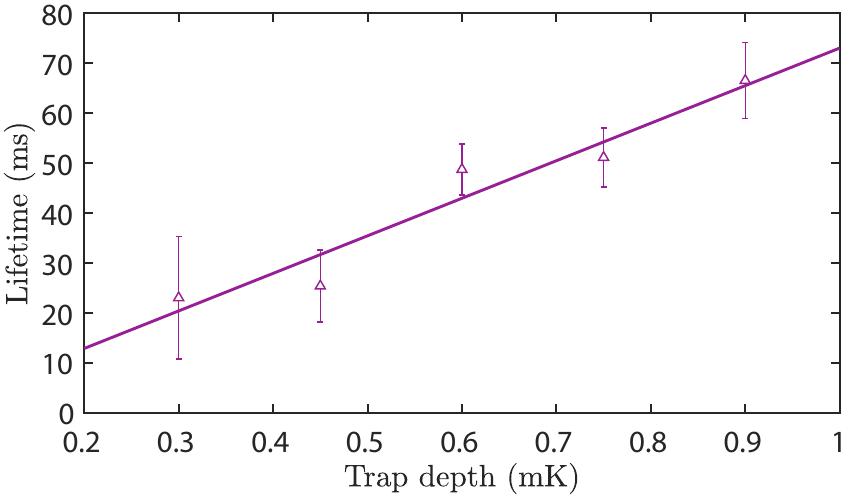}}
\caption{Atom lifetime in the standing wave dipole trap for different trap depth. The solid line is a linear fit of our measured atom lifetime for different trap depth, the fitted total heating rate from the measured data is $12.4\,\mathrm{mK/s}$.
}
\label{fig5}
\end{figure}

We also notice that the maximum transport efficiency is currently limited to about $50\%$. While nearly lossless atom transport was achieved for a transport distance of a few milimeters in free space~\cite{Schrader2001}, the reflection of the dipole trap beams on the chip surface may destroy the moving lattice antinodes in our cases. This issue might be mitigated by choosing an appropriate polarization of the dipole trap beams with incident angle satisfying the Brewster angle, since the reflection can be greatly suppressed. Finally, the intrinsic heating of atoms in the conveyor belt is tested in a static standing wave dipole trap ($\Delta\nu\mathrm{_{max}}=0$). As shown in Fig.~\ref{fig5}, a heating rate of $12.4\,\mathrm{mK/s}$ is extracted from the measurement of the trap lifetime for different trap depths, which explains the severe atomic loss when $\Delta\nu\mathrm{_{max}}<80\,\mathrm{kHz}$. 

\section{Conclusion}
We have demonstrated a hybrid photonic-atomic chip platform and successfully transported an ensemble of cold atoms from free space towards the chip with an optical conveyor belt. Our PAC platform is made of transparent materials, allowing full optical access for free space laser cooling, atom transport, and efficient coupling to on-chip photonic waveguides. The maximum transport efficiency of atoms is about $50\%$ with a transport distance of $500\,\mathrm{\mu m}$. In the future, by combining on-chip MOT laser cooling, optical transport and evanescent-field trapping of cold atoms, a more compact PAC platform is attainable through the implementation of advanced photonic structure fabrication and design strategies. Our PAC platform holds great potential for studies of atom-photon interactions and the realization of single-photon-level optical nonlinearity, which could find applications in quantum information science and quantum sensing.

\section{Acknowledgments}
This work was funded by the National Key R\&D Program (Grant No.~2021YFF0603701), the National Natural Science Foundation of China (Grants No.~U21A20433, No.~U21A6006, No.~92265210, No.~12104441, No.~12134014, No.~61905234, and No.~11974335), and the USTC Research Funds of the Double First-Class Initiative (Grant No.~YD2030002007).  CLZ was also supported by the Fundamental Research Funds for the Central Universities, and USTC Research Funds of the Double First-Class Initiative. This work was partially carried out at the USTC Center for Micro and Nanoscale Research and Fabrication.

\bibliographystyle{Zou}
\bibliography{areference}
\end{document}